\pdfoutput=1
\documentclass[sigconf]{acmart}

\usepackage{booktabs} 

\usepackage[nolist]{acronym}
\usepackage{adjustbox}
\usepackage{balance}
\usepackage[inline]{enumitem}
\usepackage{graphicx}
\usepackage{pgfplots}
\usepgfplotslibrary{statistics} 
\usepackage{xcolor}
\usepackage{url}
\usepackage{tikz}
\pgfplotsset{compat=1.16}
\usepackage{subcaption}
\usepackage{algpseudocode}
\usepackage{algorithm}

\copyrightyear{2023}
\acmYear{2023}
\setcopyright{acmcopyright}\acmConference[SAC '23]{The 38th ACM/SIGAPP Symposium on Applied Computing}{March 27-31, 2023}{Tallinn, Estonia}
\acmBooktitle{The 38th ACM/SIGAPP Symposium on Applied Computing (SAC '23), March 27-31, 2023, Tallinn, Estonia}
\acmPrice{15.00}
\acmDOI{10.1145/3555776.3577677}
\acmISBN{978-1-4503-9517-5/23/03}

\newfloat{protocol}{tbp}{lop}
\floatname{protocol}{Protocol}

\begin{document}

\title{Distributed Key Generation with Smart Contracts using zk-SNARKs}
  
\renewcommand{\shorttitle}{Distributed Key Generation with Smart Contracts using zk-SNARKs}

\author{Michael Sober}
\orcid{0000-0002-9612-9022}
\affiliation{
    \institution{Christian Doppler Laboratory for Blockchain Technologies for the Internet of Things}
}
\affiliation{%
  \institution{TU Hamburg}
  \city{Hamburg} 
  \country{Germany} 
}
\email{michael.sober@tuhh.de}

\author{Max Kobelt}
\affiliation{
    \institution{Christian Doppler Laboratory for Blockchain Technologies for the Internet of Things}
}
\affiliation{%
  \institution{TU Hamburg}
  \city{Hamburg} 
  \country{Germany}
}
\email{max.kobelt@tuhh.de}

\author{Giulia Scaffino}
\orcid{0000-0001-5680-3003}
\affiliation{
    \institution{Christian Doppler Laboratory for Blockchain Technologies for the Internet of Things}
}
\affiliation{%
  \institution{TU Wien}
  \city{Vienna} 
  \country{Austria}
}
\email{giulia.scaffino@tuwien.ac.at}

\author{Dominik Kaaser}
\orcid{0000-0002-2083-7145}
\affiliation{%
  \institution{TU Hamburg}
  \city{Hamburg} 
  \country{Germany}
}
\email{dominik.kaaser@tuhh.de}

\author{Stefan Schulte}
\orcid{0000-0001-6828-9945}
\affiliation{
    \institution{Christian Doppler Laboratory for Blockchain Technologies for the Internet of Things}
}
\affiliation{%
  \institution{TU Hamburg}
  \city{Hamburg} 
  \country{Germany}
}
\email{stefan.schulte@tuhh.de}

\renewcommand{\shortauthors}{Sober et al.}

\begin{acronym}
    \acro{EVM}{Ethereum Virtual Machine}
    \acro{ZKP}{Zero-Knowledge Proof}
    \acro{DKG}{Distributed Key Generation}
    \acro{VSS}{Verifiable Secret Sharing}
    \acro{PVSS}{Publicly Verifiable Secret Sharing}
    \acro{zk-SNARK}{Zero-Knowledge Succinct Non-interactive Arguments of Knowledge}
    \acro{zk-SNARKs}{Zero-Knowledge Succinct Non-interactive Arguments of Knowledge}
    \acro{zk-STARKs}{Zero-Knowledge Succinct (Scalable) Transparent Arguments of Knowledge}
    \acro{R1CS}{Rank-1 Constraint-System}
    \acro{QAP}{Quadratic Arithmetic Program}
    \acro{VOC}{Verifiable Off-Chain Computation}
    \acro{EVM}{Ethereum Virtual Machine}
    \acro{NIZK}{Non-interactive Zero-Knowledge Proof}
    \acro{TID}{Threshold Information Disclosure}
    \acro{CP-ABE}{Ciphertext-Policy Attribute-Based Encryption}
    \acro{PoS}{Proof of Stake}
    \acro{MPC}{Multiparty Computation}
    \acro{DSL}{Domain-Specific Language}
\end{acronym}
\begin{abstract}

\ac{DKG} is an extensively researched topic as it is fundamental to threshold cryptosystems. Emerging technologies such as blockchains benefit massively from applying threshold cryptography in consensus protocols, randomness beacons, and threshold signatures. However, blockchains and smart contracts also enable further improvements of \ac{DKG} protocols by providing a decentralized computation and communication platform.

For that reason, we propose a \ac{DKG} protocol that uses smart contracts to ensure the correct execution of the protocol, allow dynamic participation, and provide crypto-economic incentives to encourage honest behavior. The \ac{DKG} protocol uses a dispute and key derivation mechanism based on \ac{zk-SNARKs} to reduce the costs of applying smart contracts by moving the computations off-chain, where the smart contract only verifies the correctness of the computation.

\end{abstract}

\ccsdesc[500]{Security and privacy~Cryptography}
\ccsdesc[500]{Computer systems organization~Distributed architectures}

\keywords{Distributed key generation, zero-knowledge proofs, smart contracts, blockchain}

\maketitle

\acresetall
\acused{zk-SNARK}

\section{Introduction}
\label{sec:introduction}

Threshold cryptography allows for the shared usage of cryptosystems with no central authority~\cite{desmedt1994threshold}. A threshold cryptosystem distributes trust among all participants, while conventional cryptosystems often involve a trusted third party. Fundamental to threshold cryptosystems is the ability to generate distributed private keys by executing a \ac{DKG} protocol. Such a protocol allows a group of participants to generate a distributed private key where each party only possesses a share of the private key while not having any knowledge about the private key itself. \ac{DKG} protocols have been subject to ongoing research over the last two decades~(see~Section~\ref{sec:background}). Notably, the latest spike of interest in blockchain technology paved the way for new applications and improvements.

Blockchain technology is not only the underlying technology of cryptocurrencies like Bitcoin~\cite{nakamoto2008bitcoin} but has also found applications in various other fields~\cite{mollah2020blockchain,dai2019blockchain}. A blockchain is a tamper-resistant distributed ledger of transactions managed by a peer-to-peer network of nodes. The second generation of blockchains, e.g., Ethereum~\cite{wood2014ethereum}, allows the execution of so-called smart contracts, which are deterministic programs stored on the blockchain. The decentralized nature of blockchains makes them an excellent application for threshold cryptosystems, e.g., for consensus, public randomness, or threshold signatures using a \ac{DKG} protocol~\cite{kokoris2020asynchronous}.

\ac{DKG} protocols are not only important for applications in the field of blockchain technology but also the blockchain, especially smart contracts, can help to improve \ac{DKG} protocols~\cite{schindler2019ethdkg, stengele2021ethtid}. Usually, in \ac{DKG} protocols, misbehaving participants are excluded from the protocol’s further execution and not held accountable for their actions. Thus, trying to attack the protocol may seem beneficial as it does not entail serious consequences. As we will discuss in the paper at hand, a smart contract can augment a \ac{DKG} protocol with crypto-economic incentives and the possibility for dynamic participation such that participants are encouraged to execute the protocol and behave honestly.

Further, external parties need to verify that a given public key was generated according to the \ac{DKG} protocol’s specification and belongs to a specific group of participants that generated the key. While it is immediately apparent to the participants, this verification process is more difficult for parties external to the protocol. Therefore, we discuss the utilization of smart contracts to ensure the correct execution of the \ac{DKG} protocol and provide a transparent view of the actions taken by each participant.

While the application of smart contracts introduces additional overhead, there are mechanisms to keep it as low as possible. These include \acp{VOC} with the help of \acp{ZKP}, where a smart contract only needs to verify a short proof and does not need to perform all calculations on the blockchain~\cite{korbel2021blockchain}.

Therefore, we propose a \ac{DKG} protocol based on smart contracts. The smart contract allows the participants to execute the different phases of the protocol by facilitating the execution. The underlying blockchain constitutes the communication channel, whereby no direct connection between the participants is necessary. Further, the smart contract offers a lightweight dispute mechanism by enabling \ac{VOC} with \ac{zk-SNARKs}. This mechanism allows for the punishment of misbehavior such as slashing provided collateral, thus making the participants accountable for their actions. Accordingly, we provide the protocol's specification and create a prototypical implementation for \ac{EVM}-based blockchains. Further, we evaluate the costs of using the smart contract as well as the performance and memory usage for generating the proofs.

The rest of this paper is structured as follows: in Section~\ref{sec:background}, we present some background information and the underlying concepts. Then, we propose the protocol in Section~\ref{sec:system} and describe its implementation in Section~\ref{sec:implementation}. We evaluate the protocol in Section~\ref{sec:evaluation} and discuss the related work in Section~\ref{sec:related_work}. Finally, we conclude the paper in Section~\ref{sec:conclusion}.
\section{Background}
\label{sec:background}

In this section, we present the concepts behind the proposed protocol. For this, we briefly describe the most important aspects of Secret Sharing, Distributed Key Generation, and Zero-Knowledge Proofs.

\subsection{Secret Sharing}

Secret Sharing was introduced by Shamir~\cite{shamir1979share} to divide a secret $S$ into $n$ pieces such that $k$ pieces or more are required to recover $S$. Providing only $k-1$ parts, it is not possible to recover $S$ and does not leak any information about $S$. Initially, a dealer picks a polynomial $f(x) = c_0 + c_1x + \dots + c_{k-1}x^{k-1}$ of degree $k-1$ in which $c_0=S$ and chooses the remaining coefficients $c_1 \dots c_{k-1}$ randomly~\cite{katz2020introduction}.

After that, a dealer evaluates $f(x_i)$ with $x_i=1 \dots n$ over a finite field $\mathbb{F}_p$ and sends the points $(x_i, f(x_i))$ to the participants through a private channel. For $k$ distinct points, there is only one polynomial that passes through all of these points. Therefore, it is possible to use polynomial interpolation, e.g., Lagrange interpolation, to reconstruct the polynomial. With the collaboration of $k$ shareholders, the shareholders can reconstruct the polynomial and evaluate $f(0)$ to recover $S$.

Shamir's scheme has the disadvantage that the dealer is a trusted party that can distribute incorrect shares. Further, shareholders can provide invalid shares during the reconstruction of the secret. This issue led to the emergence of \ac{VSS} schemes. Ever since \ac{VSS} first has been mentioned by Chor et al.~\cite{chor1985verifiable}, many different VSS schemes followed, e.g.,~\cite{feldman1987practical, rabin1989verifiable, pedersen1991non,stadler1996publicly, schoenmakers1999simple}, which exhibit distinct characteristics. In the following discussion, we limit ourselves to non-interactive VSS schemes, more specifically to Feldman's \ac{VSS} scheme, as no additional communication between the dealer and the other participants is necessary.

In Feldman's \ac{VSS} scheme, only the dealer sends messages, while all shareholders can verify if the received share is correct without sending any messages. In addition to sending the shares to each participant, the dealer broadcasts public commitments $c_iG$ for $0 \leq i \leq k-1$ to the coefficients of the random polynomial, where $G$ is a generator of a cyclic group of elliptic curve points. Every shareholder has access to the public polynomial $F(x)$ defined as
\begin{equation*}
\label{eq:pub_poly}
    F(x) =  c_0G + c_1Gx + \dots + c_{k-1}Gx^{k-1} .
\end{equation*}
A shareholder can use this public polynomial to verify that the received share is correct by evaluating $f(x_i)G = F(x_i)$.

\subsection{Distributed Key Generation}

In threshold cryptography, there is no single party possessing all the necessary information to encrypt, decrypt or sign a message. Instead, a threshold cryptosystem requires that a certain subset of size at least $t$ out of $n$ parties need to collaborate on these actions. Unfortunately, with \ac{VSS} schemes, there is still the problem that a dealer is a central authority that knows the shared secret. A dealer can still act without collaborating with other parties if it does not discard the secret, thus leading to some centralization of trust.

\ac{DKG} enables multiple parties to collectively generate a distributed private key with no central point of trust. Each party only has a secret share of the private key, but neither party knows the complete private key, which is also never recovered. While many different \ac{DKG} protocols already exist~\cite{pedersen1991threshold, abraham2021reaching, dehez2021blockchain, das2021practical, gennaro1999secure, kate2009distributed, kate2012distributed, kokoris2020asynchronous}, we limit ourselves to Pedersen's \ac{DKG}~\cite{pedersen1991threshold} as it has been widely used to create threshold cryptosystems. Further, it forms the basis of our proposed protocol~(see~Section~\ref{sec:protocol}).

Pedersen's \ac{DKG} protocol allows a set of participants $P=\{P_1, \dots, P_n\}$ to generate a distributed private key where at least $t$ participants need to collaborate to use the private key. For that, every party $P_i$ executing the \ac{DKG} protocol shares a signed secret $s_{ij} = f_i(j)$ with all the other parties $P_j \in P \setminus P_i$ using Feldman's VSS scheme. Upon receiving an invalid share $s_{ij}$ from any other participant, a participant can issue a complaint by publishing $s_{ij}$ and the signature. More than $t$ complaints against a specific participant ensure its disqualification, i.e., the exclusion from the protocol's execution. A participant $P_j$ can compute its share of the distributed private key by calculating $s_j = \sum_{i=1}^{i=n}s_{ij}$. Finally, the participants can compute the shared public key via the previously published commitments to the random polynomials by calculating $H(x)=\sum_{i=1}^{i=n}F_i$ and evaluating $H(0)$.

\subsection{Zero-Knowledge Proofs}
\label{sub:zkps}

A \ac{ZKP} allows a prover to convince a verifier that a statement is true without revealing any information, i.e., the verifier has zero knowledge besides the validity of the statement~\cite{goldwasser1989knowledge}. Many distinct blockchains utilize \acp{ZKP} to enhance privacy or scalability~\cite{hopwood2016zcash, bonneau2020mina}. Today the most popular \ac{ZKP} systems comprise \ac{zk-SNARKs}~\cite{ben2014succinct}, \ac{zk-STARKs}~\cite{ben2019scalable}, and Bulletproofs~\cite{bunz2018bulletproofs}. With the help of these proof systems, it is possible to verify computational integrity, i.e., whether the output of a program given some public and private inputs is correct according to its specification without re-executing the whole program to verify the given output. Instead, a verifier only needs to verify a small proof~\cite{kosba2020mirage}.

The verification complexity and proof size of \ac{zk-STARKs} and Bulletproofs depend on the computational complexity of the program, while for \ac{zk-SNARKs}, it is constant. A disadvantage is that \ac{zk-SNARKs} depend on a trusted setup, whereas this dependency is non-existent for \ac{zk-STARKs} and Bulletproofs. However, the short proof size and the constant verification complexity make \ac{zk-SNARKs} an excellent candidate for the application within smart contracts. Therefore, in the following, we only focus on \ac{zk-SNARKs}.

The first practical implementation of \ac{zk-SNARKs} was the Pinocchio protocol~\cite{parno2013pinocchio}, followed by Groth16~\cite{groth2016size}, which uses pairing-friendly elliptic curves to validate the proofs. For the protocol, a program is transformed into an arithmetic circuit, then a \ac{R1CS}, and finally encoded as a \ac{QAP}. The output of a program is correct if the respective solution to the \ac{QAP} is also valid. The succinctness and zero-knowledge aspect of such proofs come from using random sampling and homomorphic encryption.

As discussed in Section~\ref{sec:client}, the proposed \ac{DKG} protocol requires executing elliptic curve operations within a circuit. Therefore, we need to clarify possible limitations which affect our solution. The limiting factor to implementing elliptic curve cryptography within a circuit is that a circuit uses modular arithmetic over a finite field $\mathbb{F}_r$. Therefore, only elliptic curves defined over $\mathbb{F}_r$ are usable within a circuit. These curves can then enable the implementation of specific cryptographic protocols~\cite{belles2021twisted}. Another possibility is to create a cycle of elliptic curves where the proof is generated using another curve and verify this proof in a circuit using $\mathbb{F}_r$.
\section{System}
\label{sec:system}

In this section, we propose a \ac{DKG} protocol that uses smart contracts for decentralized computation and communication in combination with zk-{SNARKs} to reduce its costs. We start with a brief overview and continue with an explanation of the communication model, cost model, smart contracts, and dynamic participation. After that, we explain the design of the protocol and provide an exact specification, including the different phases: \textit{Share Distribution}, \textit{Dispute}, and \textit{Key Derivation}.

The execution of the protocol starts with the \textit{Share Distribution} phase. Each participant generates a secret and distributes the shares among the other participants using Feldman's \ac{VSS} and a smart contract. Afterward, the protocol reaches the \textit{Dispute} phase, during which the participants verify the received shares and issue disputes against dealers who distributed invalid shares. The accused dealer can generate a \ac{zk-SNARK} to justify that the distributed share is correct and the dispute is unjustified. Not resolving a dispute leads to exclusion from the protocol and crypto-economic punishments. Finally, the protocol reaches the \textit{Key Derivation} phase, in which every participant uses the correctly distributed shares to derive the shared public key. For submitting the public key to the smart contract, any participant can generate a \ac{zk-SNARK} to prove that the derived public key is correct.

\subsection{Communication Model}
\label{sec:communication}

We assume a synchronous communication model in which all messages are delivered within a fixed time bound $\Delta$ and consider the blockchain a public broadcast channel where parties can send and receive messages. 

Further, instead of establishing point-to-point connections between participants, the participants use the public broadcast channel to exchange direct messages. For that, the participants generate a public/private key pair and disclose the public key to all other participants using the blockchain. The participants use the public key and some unique round-specific information to generate a shared key with its counterparts. This shared key is then used as part of a symmetric key encryption procedure to encrypt communication over the public channel.

\subsection{Cost Model}
\label{sec:cost}

Additionally, we assume that a public blockchain requires transaction fees to be paid with the blockchain's native currency. The transaction fees incentivize miners to spend their computational resources to create new blocks and execute transactions. Further, the transaction fees help to protect the platform from misuse since wasting resources also entails considerable costs.

The transaction fees depend on the actual computation steps and required storage space. As such, transactions which only transfer value between different accounts incur lower transaction fees than transactions that trigger the execution of a smart contract performing complex calculations or require a lot of storage~\cite{brandstaetter2020opt}. Especially, elliptic curve operations are very cost intensive. Therefore, our \ac{DKG} protocol uses zk-{SNARKs} to move these computations off-chain.

\subsection{Smart Contracts}
\label{sec:smart contracts}

Initially, the protocol requires a trusted setup ceremony, which is necessary for using \ac{zk-SNARKs} (see Section~\ref{sub:zkps}). This procedure generates the proving and verification keys for the circuits of Alg.~\ref{alg:justify} and \ref{alg:derive} (see Section~\ref{sec:protocol}) to off-chain the computations for justification and key derivation. The execution of the trusted setup generates a simulation trapdoor referred to as toxic waste, which needs to be discarded because it would enable the generation of fake proofs~\cite{bowe2018multi}. Therefore, it is recommended to use \ac{MPC} to distribute the trust among multiple parties~\cite{bowe2017scalable}. This approach ensures that as long as one honest party discards its part of the secret, it prevents the possibility of recovering the overall secret.

After executing the trusted setup ceremony, the smart contracts are deployed on the blockchain. The protocol requires three different smart contracts: the \textit{Justification}, \textit{Key Derivation}, and \textit{zkDKG} contracts. The \textit{Justification} and \textit{Key Derivation} contracts provide the same functionality as both contracts only have to verify \ac{zk-SNARKs} for the different circuits used during the \textit{Dispute} and \textit{Key Derivation} phases of the protocol. They only differ with respect to the inputs and verification keys. The \textit{zkDKG} contract manages the whole execution of the \ac{DKG} protocol. The participants use it to distribute the shares, issue disputes, justify their actions during disputes, and derive the public key. As justification and key derivation are based on \ac{zk-SNARKs}~(see~Sections~\ref{sec:dispute}~and~\ref{sec:derivation}), the \textit{zkDKG} contract interacts with the \textit{Justification} and \textit{Key Derivation} contracts to verify the validity of the proofs.

Further, the \textit{zkDKG} contract requires the configuration of different parameters, including the addresses of the verification contracts, threshold, time limits of the different phases, curve parameters, and others that might be necessary to dynamically form the set of participants.

\subsection{Dynamic Participation}
\label{sec:dynamic_participation}

Finally, after finishing the deployment of the smart contracts, different parties can start the formation process to qualify as participants. In contrast to conventional \ac{DKG} protocols, which do not use a blockchain, several options are available to dynamically form the set of participants, where we do not want to limit ourselves to a specific one. Here, the \textit{zkDKG} contract defines and strictly enforces who is allowed to participate in the execution of the protocol.

Among other things, one can imagine a permissioned setting in which the \textit{zkDKG} contract defines that only its owner can decide who is allowed to join. The owner only has to provide a predefined list of the selected parties to the smart contract. The \textit{zkDKG} contract checks if the potential party is allowed to join and only after successful authentication adds it to the current set of participants. The problem here is that the owner represents a central authority. 

However, other permissionless mechanisms potentially allow everyone to join without having a central authority. The \textit{zkDKG} contract could implement a naive mechanism to allow everyone to join until reaching a certain number of participants as long as they deposit some tokens as collateral. However, this approach is impractical in most cases since it does not provide any Sybil resistance, which would allow anyone to register with multiple identities~\cite{douceur2002sybil}. In the worst case, a single party could register enough identities to gain sole control and knowledge of the generated key.

An alternative solution is to let the potential participants stake tokens not only as collateral but also to let the number of tokens decide who is allowed to participate, as is done in systems based on \ac{PoS}~\cite{kiayias2017stake}. Following this direction, the \textit{zkDKG} contract specifies, e.g., that only the $n$ parties with the highest deposited stake may participate. Should participants misbehave during the execution of the protocol, they might get penalized by slashing their stake. This approach encourages participants to behave honestly and improves Sybil resistance by encouraging the creation of only one identity staking enough tokens. Another modification of this approach would also allow regular token holders to delegate their tokens to potential candidates to elect them as participants.

\subsection{Protocol}
\label{sec:protocol}

After clarifying the necessary steps prior to the execution of the protocol, we continue with its description and specification~(see~Protocol~\ref{prot:dkg}). For that, we examine the \textit{Share Distribution}, \textit{Dispute}, and \textit{Key Derivation} phases and describe the different execution steps sequentially.

\subsubsection{Share Distribution}
\label{sec:share}

\begin{protocol}

    A set of participants $P=\{P_1, \dots, P_n\}$ generates a distributed private key with threshold $t$ using a $zkDKG$ smart contract.

    \caption{Distributed Key Generation Protocol}
    \label{prot:dkg}

    \begin{enumerate}
        \item \textbf{Share Distribution} \label{protocol:share}
        \begin{enumerate}
            \item A participant $P_i \in P$ retrieves the threshold computed by the $zkDKG$ contract.
            \item $P_i$ generates a random polynomial $f_i(x)$ of degree $t-1$ and its corresponding public polynomial $F_i(x)$ with coefficients $C_i = \{c^i_0G,...,c^i_{t-1}G\}$.
            \item $P_i$ computes the set of shares $S_i = \{ f_i(j) \, | \, 0 < j < n, j \neq i \}$ and encrypts each share $s_{ij} \in S_i$.
            \item $P_i$ calls the \textit{distribute} function of the $zkDKG$ contract and provides $C_i$ and $S_i$. The $zkDKG$ contract then verifies that:
            \begin{enumerate}
                \item the distribution phase is ongoing,
                \item $P_i$ has not already distributed $S_i$,
                \item provided $|S_i| = |P \setminus P_i|$ shares,
                \item $|C_i| = t-1$ coefficients
            \end{enumerate} and otherwise reverts.
            \item The $zkDKG$ contract stores $c^i_0G$, $hash(C_i)$, and $hash(S_i)$ and notifies $P$ that $P_i$ submitted a share.
            \item A participant $P_j \in P$ verifies the received share $s_{ij}$ by checking that $F_i(j)= f_i(j)G = s_{ij}G$. If the share is invalid, $P_j$ continues with Step~\ref{protocol:dispute} and otherwise with Step~\ref{protocol:derive}.
        \end{enumerate}
        
        \item \textbf{Dispute} (optional) \label{protocol:dispute}

        \begin{enumerate}
            \item $P_j$ calls the \textit{dispute} function of the $zkDKG$ contract providing the index $i$ of the disputee $P_i$ and $S_i$. The $zkDKG$ contract then verifies that:
            \begin{enumerate}
                \item the dispute phase is ongoing,
                \item $P_i$ has not already been disputed,
                \item $S_i$ matches the stored $hash(S_i)$
            \end{enumerate} and otherwise reverts.
        \item The $zkDKG$ contract notifies $P_i$ about the dispute including $i$, $j$, $S_i$, and the time limit $\tau$ to justify.
        \item $P_i$ generates a zk-SNARK $\pi_J$ for the execution of Alg.~\ref{alg:justify} and uses the \textit{justify} function of the $zkDKG$ contract to submit $\pi_J$.
        \item The $zkDKG$ contract validates $\pi_J$ and if $\pi_J$ is valid resolves the dispute. In case $\pi_J$ is invalid, the dispute remains unresolved.
        \end{enumerate}
        
        \item \textbf{Key Derivation} \label{protocol:derive}
        \begin{enumerate}
            \item The $zkDKG$ contract verifies that the key derivation phase is ongoing.
            \item The $zkDKG$ contract checks for expired disputes and sets the coefficients of excluded participants to zero.
            \item $P_i$ derives the public key $pk$ by calculating $H(x)=\sum_{j=1}^{j=n}F_j$ and evaluating $H(0)$.
            \item Any participant generates a zk-SNARK $\pi_R$ for the execution of Alg.~\ref{alg:derive} and uses the \textit{derive} function of the $zkDKG$ contract to submit $pk$ and $\pi_R$.
            \item The $zkDKG$ contract verifies $\pi_R$, notifies $P$ about the successful submission, and stores $pk$. If $\pi_R$ is invalid the \textit{zkDKG} contract reverts.
            \item The $zkDKG$ contract slashes disqualified participants.
        \end{enumerate}
        
    \end{enumerate}
\end{protocol}

After determining the participants, the actual execution of the protocol starts with the \textit{Share Distribution} phase~(see~Step~\ref{protocol:share} of Protocol~\ref{prot:dkg}) in which every participant acts as a dealer in one of $n$ parallel executions of Feldman's \ac{VSS}. For that, a participant $P_i$ has to retrieve the targeted threshold $t$ from the \textit{zkDKG} contract. The next step is to generate the random polynomial $f_i(x)$ of degree $t-1$ and commit to the polynomial's coefficients to create its public counterpart $F_i(x)$ with coefficients $C_i = \{c^i_0G,...,c^i_{t-1}G\}$. After that, a participant generates its shares $S_i$ by computing $f_i(j)$ for every other participant $P_j \in P \setminus P_i$. Each share $s_{ij} \in S_i$ gets encrypted using a symmetric encryption scheme (e.g., the One-Time-Pad~\cite{schindler2019ethdkg}). For that, a participant retrieves the public keys of the other participants from the \textit{zkDKG} contract to compute a shared key with each participant using the Diffie-Hellman key exchange. Additionally, to ensure that each shared key is unique, as it is necessary for the One-Time-Pad, we concatenate it with $c^i_0G$ and take its hash as the shared key.

Following, a participant only has to broadcast its key shares $S_i$ and the coefficients $C_i$ using the blockchain. For that, the \textit{zkDKG} contract offers a \textit{distributeShares} function which is only callable for registered participants during the \textit{Share Distribution} phase of the protocol. The \textit{zkDKG} contract also ensures that the participants can only submit the right number of shares and commitments. After these checks, the \textit{zkDKG} contract stores $c^i_0G$ and only the hashes of $S_i$ and $C_i$ to save storage costs. Further, the \textit{zkDKG} contract either notifies the participants, or the participants have to poll the \textit{zkDKG} contract to check the current status. After that, a receiving participant $P_j$ can query the blockchain to get their share $s_{ij} \in S_i$, decrypt it, and verify that $F_i(j)= f_i(j)G = s_{ij}G$. The \textit{Share Distribution} phase ends after reaching the timeout or if all participants distributed their shares.

\subsubsection{Dispute}
\label{sec:dispute}

After the distribution of the shares, the \textit{zkDKG} contract reaches the \textit{Dispute} phase~(see~Step~\ref{protocol:dispute} of Protocol~\ref{prot:dkg}) of the protocol. If a participant receives an invalid share, it can issue a dispute against the dealer, which then has to justify by proving the correctness of the distributed share.  A participant issues disputes by calling the \textit{zkDKG} contract’s \textit{dispute} function and providing the index of the dealer and the shares. The \textit{zkDKG} contract verifies that it reached the \textit{Dispute} phase, checks if the hash of the provided shares matches the stored hash and that no dispute concerning this dealer is ongoing. Afterward, the \textit{zkDKG} contract notifies the dealer or respectively the dealer has to check if there are any ongoing disputes to resolve within a certain time limit. This time limit is extended with every dispute to provide the dealer enough time to justify.

This approach requires an additional interaction as the disputer first issues a dispute, and the respective dealer then has to justify its actions. By having the disputer generate a \ac{zk-SNARK} and proving that the share is invalid, it is possible to avoid additional interactions with the dealer. However, the smart contract then needs to verify the inputs by the dealer during the \textit{Share Distribution} phase to ensure that the receiver can generate a \ac{zk-SNARK} which is only possible with correct input values. Therefore, it is better to leave it to the dealer, who has the incentive to provide correct inputs, because otherwise, the dealer cannot generate a valid \ac{zk-SNARK}. This decreases the costs since there is no need for the smart contract to verify the inputs.

The dealer continues with the generation of the \ac{zk-SNARK} for Alg.~\ref{alg:justify} to prove the correctness of the disputed share. For that, the dealer has to provide its commitments $c$, secret key $sk$, public key $pk$, the public key of the disputer $pk_d$, index $i$, encrypted share $s_{enc}$, and the hash of the public parameters $h$. The costs of verifying the \ac{zk-SNARK} increase with the number of public input parameters. Therefore, we declare all public parameters as private and only pass $h$ to verify them within the circuit by comparing the hashes. Further, the dealer has to provide the commitments as uncompressed points as decompressing is too complex within the circuit.

The program~(see~Alg.~\ref{alg:justify}) for the justification of a disputed share starts by checking whether the dealer knows the secret key to the provided public key. After that, the program needs to verify that the dealer provided the correct public input parameters. For that, it hashes the commitments and subsequently hashes the result again with the remaining public input parameters. The hash is then compared with the hash provided by the dealer and only if both hashes match, it is possible to continue with the computation. Since the input parameters are correct, the program can derive the shared key from $sk$, $pk_d$, and $c$, to get the decrypted share $s_{dec}$. Finally, it checks if the public polynomial evaluates to $s_{dec}G$ and returns the result.

After computing the proof, the dealer calls the \textit{justify} function of the \textit{zkDKG} contract providing the previously computed proof. Initially, the \textit{zkDKG} contract checks if there is a dispute to resolve. After that, the \textit{zkDKG} contract computes the hash of the public input parameters and calls the \textit{Justification} contract. The \textit{zkDKG} contract reverts if the proof is invalid, and otherwise, it deletes the dispute and excludes the disputer for the unjustified dispute.

\begin{algorithm}[t]
\caption{Compute the justification that a distributed share is correct and was wrongly disputed}
\label{alg:justify}
\begin{algorithmic}[1]

\Function{justify}{$c$, $sk$, $pk$, $pk_d$, $i$, $s_{enc}$, $h$}
\State $assert(mult(sk, G) == pk)$
\State $h_c \gets hash(compress(c))$
\State $h_c \gets hash(h_c, pk, pk_d, i, s_{enc})$
\State $assert(h == h_c)$
\State $s_{dec} \gets s_{enc} - sharedKey(sk, pk_d, c)$ 
\State $actual \gets mult(s_{dec},G)$
\State $expected \gets evalPubPoly(i, c)$
\State \Return $actual == expected$
\EndFunction

\Function{sharedkey}{sk, pk, c}
\State $k \gets scalarMult(sk,pk)$
\State \Return $hash(k, c[0])$
\EndFunction

\Function{evalpubpoly}{i, c}
\State $result \gets Infinity$
    \For{$k \gets 1$ to $len(c)$}                    
        \State $result \gets mult(result, i)$
        \State $result \gets add(c[len(c)-1-k]$
    \EndFor
    \State \Return $result$
\EndFunction

\end{algorithmic}
\end{algorithm}

\subsubsection{Key Derivation}

After reaching the time limit and allowing all participants to verify their received shares and file a dispute, the final \textit{Key Derivation} phase~(see~Step~\ref{protocol:derive} of Protocol~\ref{prot:dkg}) of the protocol begins. A participant computes its overall share of the distributed private key by summing up the shares received from qualified participants, i.e., participants with no unresolved disputes. The derivation of the public key is very cost-intensive using a smart contract. Therefore, any participant can generate a \ac{zk-SNARK} for Alg.~\ref{alg:derive} to derive the public key off-chain and only submit the public key with the proof to the \textit{zkDKG} contract. A participant can derive the public key by computing $pk=\sum_{j=1}^{j=n}c_0^jG$. However, generating a proof of this computation also requires the hash of the public input parameters, i.e., the first public coefficients, to verify that a participant uses the correct input parameters.

After generating the proof, a participant calls the \textit{derive} function of the \textit{zkDKG} contract providing the public key and the proof. The \textit{zkDKG} contract checks if it is in the \textit{Key Derivation} phase and if participants have any unresolved disputes. If a participant has unresolved disputes, it is excluded by setting its first public coefficient to 0, i.e., the point at infinity. Further, the \textit{zkDKG} contract aborts the execution of the protocol if the number of excluded participants exceeds the limit specified during deployment. In the worst case, the number of participants and the threshold are equal, meaning that a single failure is enough to make the distributed private key unusable. This case requires a restart of the protocol with an updated set of participants, i.e., misbehaving participants get replaced and lose their collateral.

After that, the \textit{zkDKG} contract computes the hash of all the stored first coefficients and the submitted public key. Here also, only providing the hash of the public inputs reduces the verification costs. Then, the \textit{zkDKG} contract calls the \textit{Key Derivation} contract, providing the previously computed hash and the public key. The \textit{zkDKG} contract stores the public key if the proof is valid and otherwise reverts. Another possibility is to let the \textit{zkDKG} contract optimistically accept the public key and introduce another dispute round after the submission. This approach reduces the costs of submitting the public key and the necessary resources since the generation and verification of the proof are only required if the submitted public key is invalid. However, another dispute round introduces further complexity.

Finally, the \textit{zkDKG} contract allows multiple runs such that only a single deployment is necessary to reduce the overall costs. It is up to the creator of the \textit{zkDKG} contract to specify when to execute a reset.

\label{sec:derivation}

\begin{algorithm}[t]
\caption{Derive the distributed public key}
\label{alg:derive}
\begin{algorithmic}[1]

\Function{derive}{$coef$, $h$}
\State $h_c \gets hash(compress(coef))$
\State $assert(h == h_c)$
\State $pk \gets Infinity$
    \For{$k \gets 1$ to $len(coef)$}                    
        \State $pk \gets mult(pk, coef[i])$
    \EndFor
\State \Return $pk$
\EndFunction

\end{algorithmic}
\end{algorithm}
\section{Implementation}
\label{sec:implementation}

In this section, we describe the implementation of the client software and the smart contracts available as open-source software on GitHub\footnote{\url{https://github.com/soberm/zkDKG}}.

\subsection{Smart Contracts}
\label{sec:contracts}

We implemented the smart contracts in Solidity for the Ethereum smart contract and decentralized application platform as it is today the most popular smart contract platform with a wide range of tools and support~\cite{belotti2019vademecum}. Further, Ethereum offers precompiled contracts for elliptic curve addition, scalar multiplication, and pairing checks on the \textit{alt\_bn128} curve, which is necessary to verify the \ac{zk-SNARKs}. Since the implemented prototype works for Ethereum, it also supports other Ethereum-based blockchains out of the box. However, the solution can also be implemented for other blockchains as long as these platforms provide the necessary smart contract capabilities and verification of \ac{zk-SNARKs}.

We decided to implement the naive registration mechanism, which has already been discussed in Section~\ref{sec:dynamic_participation}, as it is sufficient for the first prototypical implementation and allowed us to focus on the more important parts of the protocol. The \textit{zkDKG} contract only allows participants to join until reaching a certain number and requires them to provide collateral. The participants can withdraw their placed collateral from the \textit{zkDKG} contract after the execution of the protocol if they behaved honestly.

The used threshold depends on the number of registered participants $n$. The \textit{zkDKG} contract computes $\lceil (n+1)/2 \rceil$, specifying that the reconstruction of the distributed private key requires more than half of the registered participants. Further, we define that at least $\lceil (2n+1)/3 \rceil$ need to successfully share their secrets to enable the submission of the public key to ensure that the system continues to work if participants misbehave. These are usually the minimum requirements considering the related work~(see~Section~\ref{sec:related_work}).

The \textit{zkDKG} contract has to store the distributed shares and commitments of the participants. Storage space on the blockchain is one of the biggest cost factors. Therefore, the \textit{zkDKG} contract keeps the storage as low as possible. For that, the \textit{zkDKG} contract only stores the first coefficients of each participant directly in the smart contract's storage while leaving the shares and commitments as part of the \textit{calldata} and only storing the Keccak256 hash, which is considerably cheaper. Unfortunately, Ethereum does not provide a cost-efficient implementation of arithmetization-oriented hash functions such as MiMC~\cite{albrecht2016mimc}, Poseidon~\cite{grassi2021poseidon}, or Rescue-Prime~\cite{szepieniec2020rescue}, which would make computing the hash inside a circuit more efficient. The \textit{zkDKG} contract emits an event to notify all participants about the recent distribution from a specific participant.

The \textit{Justification} and \textit{Key Derivation} contracts required no manual implementation. For these functionalities, the ZoKrates toolbox~(see~Section~\ref{sec:client}) provides the possibility to generate the verification contracts for the respective circuits. These generated contracts use the precompiled contracts provided by Ethereum to ensure the cost-efficient verification of the proofs.

\subsection{Client}
\label{sec:client}

In addition to the smart contracts, we also provide a prototypical implementation of the client software written in the Go programming language. The client software is the off-chain component that uses the \textit{zkDKG} contract to generate the distributed private key. For generating the \ac{zk-SNARKs}, we use the ZoKrates~\cite{eberhardt2018ZoKrates} toolbox, which enables \ac{zk-SNARKs} on Ethereum. ZoKrates offers a \ac{DSL} to create off-chain programs and compile them to arithmetic circuits. After that, a client can use ZoKrates to compute the witness for a specific circuit and generate a proof of computation using the selected proving scheme. We selected Groth16 as the proofs only consist of three elliptic curve points, which makes the verification on the blockchain cheaper.

We used the \ac{DSL} provided by ZoKrates to implement Alg.~\ref{alg:justify} and \ref{alg:derive}. The standard library of ZoKrates already provides the implementation of elliptic curve operations over the Baby Jubjub curve~\cite{belles2021twisted}. Hence, our prototype uses this curve for elliptic curve cryptography. However, the protocol supports any other curve as long as its base field matches the scalar field of the curve used for the \ac{zk-SNARKs}.

The client software uses Go language bindings for Ethereum smart contracts to avoid boilerplate code and directly interact with the smart contracts through their interfaces. Unfortunately, ZoKrates does not provide Go bindings. Therefore, the client uses Docker to spin up new containers using the ZoKrates Docker image and execute the respective commands to compute the witness and generate the proofs for Alg.~\ref{alg:justify} and \ref{alg:derive} within a container. The client retrieves the result from the container to submit the proof to the \textit{zkDKG} contract when executing the \textit{justify} or \textit{derive} functions.
\section{Evaluation}
\label{sec:evaluation}

After describing the implementation of the protocol, we use this section to evaluate the protocol concerning costs, performance, and memory. The results give insight into the overhead introduced by smart contracts and the necessary resources needed to generate the proofs for justification and key derivation.

We execute the protocol ten times each before doubling the number of participants until we reach 256 participants resulting in 70 protocol runs. During the executions, we measure the gas consumption, proof generation time, and memory usage. We perform the experiments on an Amazon EC2 instance equipped with an Intel(R) Xeon(R) Platinum 8259CL running Ubuntu 22.04 using eight cores with a clock speed of 2.50GHz. The machine has access to 32GB DDR4 RAM with a frequency of 3200MHz and an SSD with a throughput of 128 MB/s max. Further, we spin up a local blockchain instance using Hardhat network 2.11.1 to deploy and run the smart contracts.

\subsection{Costs}

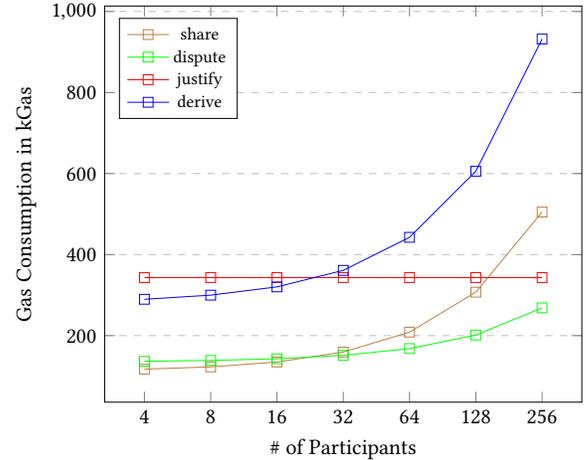
\begin{figure}[t]
\centering
\begin{adjustbox}{width=0.9\linewidth}
\begin{tikzpicture}
\begin{axis}[
    xmode=log,
    log basis x={2},
    xlabel={\# of Participants},
    ylabel={Gas Consumption in kGas},
    xtick={4,8,16,32,64,128,256},
    xticklabels={4,8,16,32,64,128,256},
    legend pos=north west,
    ymajorgrids=true,
    legend style={nodes={scale=0.8}},
    grid style=dashed
]

\addplot[
    color=brown,
    mark=square,
    ]
    coordinates {
    (4,117.227)(8,122.855)(16,134.912)(32,159.405)(64,208.617)(128,307.237)(256,504.964)
    };
    \addlegendentry{share}

\addplot[
    color=green,
    mark=square,
    ]
    coordinates {
    (4,136.646)(8,138.737)(16,142.918)(32,151.275)(64,168.004)(128,201.454)(256,268.455)
    };
    \addlegendentry{dispute}
    
\addplot[
    color=red,
    mark=square,
    ]
    coordinates {
    (4,343.211)(8,343.211)(16,343.215)(32,343.212)(64,343.219)(128,343.210)(256,343.219)
    };
    \addlegendentry{justify}
    
\addplot[
    color=blue,
    mark=square,
    ]
    coordinates {
    (4,289.694)(8,299.871)(16,320.263)(32,361.036)(64,442.584)(128,605.677)(256,931.937)
    };
    \addlegendentry{derive}
    
\end{axis}
\end{tikzpicture}
\end{adjustbox}
\caption{Gas consumption of the different phases}
\label{fig:costs}
\end{figure}

The execution of smart contracts on the Ethereum platform requires gas to protect the platform by preventing the unnecessary consumption of resources. Further, it serves as a reward for miners and incentivizes them to provide their resources. However, this is not only an inherent characteristic of the Ethereum platform but also of other smart contract platforms that need to compensate miners. Therefore, we examine the costs incurred by the protocol through executing smart contracts. We specifically look at the gas consumption of the \textit{share}, \textit{dispute}, \textit{justify} and \textit{derive} functions.

The results~(see~Fig.~\ref{fig:costs}) show that the \textit{share} and \textit{dispute} functions consume the least amount of gas. The gas consumption of the \textit{share} function averages 117 kGas for 4 participants reaching 505~kGas for 256 participants. These costs are relatively low since the shares and commitments are part of the calldata, while the \textit{zkDKG} contract only stores the respective hash values. The \textit{dispute} function consumes even less gas since only the number of shares impacts its gas consumption. Here we get 137 kGas for 4 participants and 269~kGas for 256 participants. However, the gas consumption of both functions increases linearly with the number of participants that execute the protocol.

Further, we can see that the gas consumption of the \textit{justify} function is constant at 343 kGas respecting the number of participants. The reason is that the \textit{justify} function only expects the \ac{zk-SNARK} consisting of three elliptic curve points as input. The \textit{zkDKG} contract does not execute any operations based on the number of participants and removes the dispute upon successful proof verification. However, the gas consumption increases slightly if multiple disputes are stored since the smart contract needs to loop over all inputs to find the respective index.

On the contrary, executing the \textit{derive} function consumes more gas with an increasing number of participants since the \textit{zkDKG} contract needs to compute the hash of all first coefficients to verify the \ac{zk-SNARK}. The results show that executing the \textit{derive} function consumes 290 kGas for 4 participants up to 932 kGas for 256 participants. The number of expired disputed also increases the gas consumption since the smart contract has to exclude the nodes before deriving the public key. However, since disputes are an exception the impact on gas consumption is low.

\subsection{Performance}

Not only do the smart contracts introduce additional overhead but also the generation of \ac{zk-SNARKs} for executing the \textit{justify} and \textit{derive} functions. Although this approach saves costs when executing the smart contracts, all participants have to use more of their local resources to generate the proofs. An interesting aspect here is how much time a participant spends on generating these proofs with an increasing number of participants.

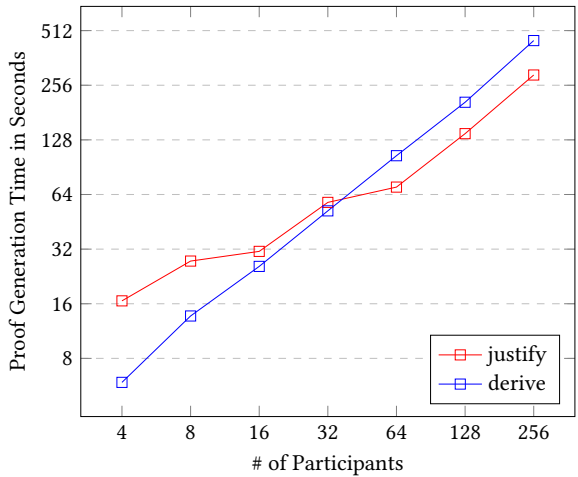
\begin{figure}[t]
\centering
\begin{adjustbox}{width=0.9\linewidth}
\begin{tikzpicture}
\begin{axis}[
    xmode=log,
    ymode=log,
    log basis x={2},
    log basis y={2},
    xlabel={\# of Participants},
    ylabel={Proof Generation Time in Seconds},
    xtick={4,8,16,32,64,128,256},
    xticklabels={4,8,16,32,64,128,256},
    ytick={8,16,32,64,128,256,512},
    yticklabels={8,16,32,64,128,256,512},
    legend pos=south east,
    ymajorgrids=true,
    grid style=dashed
]

\addplot[
    color=red,
    mark=square,
    ]
    coordinates {
    (4,16.6)(8,27.5)(16,31.1)(32,57.9)(64,70.3)(128,138.5)(256,291.5)
    };
    \addlegendentry{justify}

\addplot[
    color=blue,
    mark=square,
    ]
    coordinates {
    (4,5.9)(8,13.7)(16,25.7)(32,52)(64,104.7)(128,206.3)(256,451.2)
    };
    \addlegendentry{derive}
    
\end{axis}
\end{tikzpicture}
\end{adjustbox}
\caption{Time to generate the proofs}
\label{fig:time}
\end{figure}

The results of the executed experiments~(see~Fig.~\ref{fig:time}) show that proof generation time for justification and key derivation linearly depends on the number of participants. The proof generation time for key derivation gets higher and grows faster than for justification. The difference is that the former directly depends on the number of participants, while the latter depends on the threshold. The generation of the proofs with 4 participants takes 17 and 6 seconds, respectively. However, it already takes 292 and 451 seconds with 256 participants.

The proof generation time is usually not a large concern since blockchains are not suitable for time-critical applications. However,  configuring the timeouts of the \textit{zkDKG} contract requires attention to ensure that each participant has enough time to generate the proofs, which is particularly important for justification. Further, there is also a lot of optimization potential to reduce the proof generation time by using arithmetization-oriented hash functions and enabling multi-threading for ZoKrates.

\subsection{Memory}

Additionally, we investigate memory consumption during proof generation. Memory consumption is usually a bigger concern than proof generation time since memory consumption tends to be very high, considering that certain computations (e.g., keccak256) are not very efficient with \ac{zk-SNARKs}. Here, we also examine how the number of participants influences memory consumption.

\begin{figure}[t]
\centering
\begin{adjustbox}{width=0.9\linewidth}
\begin{tikzpicture}
\begin{axis}[
    xmode=log,
    ymode=log,
    log basis x={2},
    log basis y={2},
    xlabel={\# of Participants},
    ylabel={Memory in MB},
    xtick={4,8,16,32,64,128,256},
    xticklabels={4,8,16,32,64,128,256},
    ytick={512,1024,2048,4096,8192,16384,32768},
    yticklabels={512,1024,2048,4096,8192,16384,32768},
    legend pos=south east,
    ymajorgrids=true,
    grid style=dashed
]

\addplot[
    color=red,
    mark=square,
    ]
    coordinates {
    (4,1394)(8,1941)(16,2447)(32,3687)(64,5329)(128,9930.7)(256,21173.45)
    };
    \addlegendentry{justify}

\addplot[
    color=blue,
    mark=square,
    ]
    coordinates {
    (4,451.5)(8,908.3)(16,1790)(32,3582)(64,7180.9)(128,14061.2)(256,27708.9)
    };
    \addlegendentry{derive}
    
\end{axis}
\end{tikzpicture}
\end{adjustbox}
\caption{Memory usage during proof generation}
\label{fig:memory}
\end{figure}
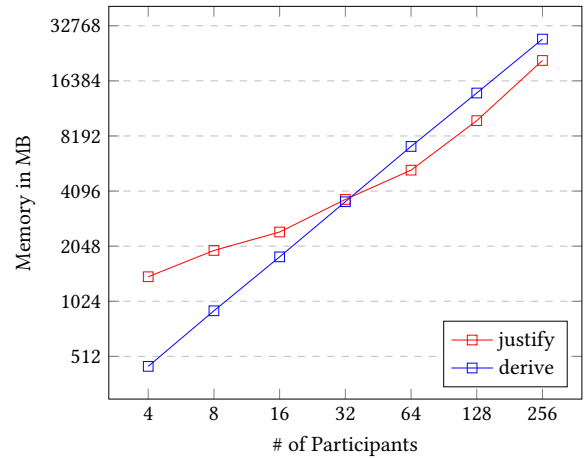

The measurements of memory consumption~(see~Fig.~\ref{fig:memory}) show that it also increases linearly with the number of participants. The memory consumption behaves the same as the proof generation time. The generation of the proof for justification consumes 1.39 GB for 4 participants up to 21.2 GB for 256 participants. For key derivation, we get a memory consumption of 452 MB for 4 participants and 27.7 GB for 256 participants. Therefore, we get a higher entry barrier for the participants since they need the resources to ensure they can generate the proofs. However, there is also potential to optimize memory consumption by using arithmetization-oriented hash functions.

\section{Related Work}
\label{sec:related_work}

\ac{DKG} protocols have always received a lot of attention in research, as they constitute the primary building block of threshold cryptosystems. Although blockchain technology already benefits from the application of threshold cryptography, only a few works, to the best of our knowledge, explore the application of smart contracts to enhance the protocols themselves.

In 1991, Pedersen~\cite{pedersen1991threshold} introduced the first \ac{DKG} protocol, which allows multiple participants to generate a shared secret, where each party only possesses a share of the secret and has no further information about the shared secret. Every participant acts as a dealer in one of $n$ parallel executions of Feldman's \ac{VSS} scheme (see~Section~\ref{sec:background}). In later work, Gennaro et al.~\cite{gennaro1999secure} propose the Joint-Feldman \ac{DKG} protocol, which builds upon Pedersen's \ac{DKG} protocol but also ensures the uniform distribution of the shared secret. Pedersen's \ac{DKG} protocol provides the foundation for our proposed protocol. While it does not guarantee uniform randomness of the shared secret, it remains secure, and only specific use cases require this property~\cite{gennaro2002revisiting, gennaro2003secure}.

Kate and Goldberg~\cite{kate2009distributed} propose the first \ac{DKG} protocol, which works in an asynchronous network setting. For that, the authors define an asynchronous \ac{VSS} scheme adopting a hybrid model for a network with $n > 3t + 2f + 1$ nodes, where an adversary may control $t$ nodes, and $f$ nodes can fail. The \ac{VSS} protocol includes a sharing and reconstruction protocol and a recovery mechanism. The work shows that asynchronous \ac{DKG} protocols require a Byzantine agreement scheme. Therefore, the authors use a leader-based agreement protocol in combination with the previously specified \ac{VSS} scheme to create a \ac{DKG} protocol. In subsequent work, Kate et al.~\cite{kate2012distributed} provide another protocol version to guarantee uniform randomness of the shared secret. Further, they provide an implementation of the protocol and analyze its performance.

In~\cite{kokoris2020asynchronous}, the authors present another asynchronous DKG protocol. However, the protocol enables the generation of a shared secret with a dual $(f, 2f + 1)$ threshold with $f$ faulty participants. The authors introduce an asynchronous high-threshold \ac{VSS} scheme that separates the reconstruction and recovery thresholds using an asymmetric bivariate polynomial. This \ac{VSS} scheme forms the basis for a weak \ac{DKG} protocol which eventually leads to a common key. By adding another mechanism to produce common randomness and implementing an agreement protocol, the authors were able to create the final asynchronous \ac{DKG} protocol. Following a different approach, Abraham et al.~\cite{abraham2021reaching} present a more efficient solution improving on the results presented in~\cite{kokoris2020asynchronous}. Further, a work by Das et al.~\cite{das2021practical} also promises improvements while providing compatibility with current threshold cryptosystems.

Schindler et al.~\cite{schindler2019ethdkg} present a \ac{DKG} protocol that uses Ethereum as its communication and verification platform. The protocol is based on the Joint-Feldman \ac{DKG} protocol~\cite{gennaro1999secure} and incorporates further improvements from \cite{neji2016distributed} to address biasing attacks. The application of blockchain technology allows for the dynamic selection of participants and the augmentation of the protocol with crypto-economic incentives. Similar to our protocol~(see~Section~\ref{sec:protocol}), it uses Feldman's \ac{VSS} to create and distribute the shares through the blockchain. However, there are some differences regarding the management of disputes. The authors describe a dispute mechanism where shareholders have to publish an invalid share and the key to decrypt the share. A disputer has to provide a \ac{NIZK} to show the equality of two discrete logarithms to ensure the correctness of the published key. After that, a smart contract can then decrypt and verify the share. In our protocol, we stick to the notion that a dealer needs to prove that it behaved correctly instead of letting the shareholder prove that the dealer violated the protocol. Further, our protocol requires that the dealer submits a \acs{zk-SNARK} to prove that it behaved correctly, which reduces the on-chain computations to a bare minimum. Despite these differences, the approach by Schindler et al. comes closest to the work at hand. 

In~\cite{stengele2021ethtid}, the authors present a smart contract for \ac{TID} based upon the work in~\cite{schindler2019ethdkg}. This approach allows users to disclose certain pieces of information simultaneously such that all or no users disclose their information. The main building block of the \ac{TID} mechanism is a \ac{DKG} protocol that allows for the scheduled reconstruction of the private key. The authors extended and simplified the DKG protocol by Schindler et al.~\cite{schindler2019ethdkg} to generate a threshold ElGamal key pair for encryption and decryption, ultimately making it more gas-efficient for the \ac{TID} use case. After $t$ or more participants have contributed to restoring the private key, any participant can submit it to the smart contract. The smart contract enables the disclosure of the information to the public by providing the information and the submitted private key. While the authors were able to optimize the \ac{DKG} protocol, it is only applicable to specific use cases.

The authors of~\cite{dehez2021blockchain} propose a blockchain-based \ac{DKG} protocol that ensures privacy and traceability. The protocol utilizes a blockchain, e.g., Bitcoin, with a pegged sidechain to improve scalability. Here, the main blockchain only acts as the coordinator and enables the anonymization of participants through a coin mixing service. The participants lock their coins on the main blockchain, generate the shares using the protocol from \cite{gennaro2003secure} and execute the sharing phase on the pegged sidechain. Participants can also issue dispute claims on the sidechain in case another participant violates the rules of the protocol. Finally, the participants can unlock their coins after successfully executing the protocol. While this solution provides better scalability through the pegged sidechain, it also introduces additional complexity.

Zhang et al.~\cite{zhang20221} present a publicly verifiable \ac{DKG} protocol that uses Ethereum for trustless computation and communication. The authors explore the application of \ac{CP-ABE} instead of \ac{PVSS} to lower the communication and computation overhead. Similar to \ac{PVSS}, \ac{CP-ABE} also enables the participants to create a shared publicly verifiable secret. The protocol only needs a single round as the smart contracts directly verify the proofs, thus removing the dispute phase. Unfortunately, the protocol incurs very high verification costs.

While our solution also entails a cost, time, and memory overhead, it is more cost-efficient than the current solutions based on blockchain technology since we move the expensive computations off-chain. Further, the availability of additional precompiled contracts to enable complex cryptographic operations besides the verification of the proof does not restrict our solution since there is no need for the smart contract to execute these complex calculations on the blockchain.
\section{Conclusion}
\label{sec:conclusion}

Threshold cryptography is an essential building block for creating threshold cryptosystems used by various applications, especially in the blockchain field.  In this work, we showed that blockchains also open new possibilities which help to advance threshold cryptosystems by offering a platform for decentralized computation and communication.

For this, we introduced a \ac{DKG} generation protocol that uses the potential of integrating blockchain technology into the protocol. Smart contracts enable  the dynamic participation of participants, augment the protocol with crypto-economic incentives to establish accountability, ensure the correct execution of the protocol and provide a public broadcast channel through the blockchain. The protocol enables participants to use \ac{zk-SNARKs} for justification and key derivation to execute heavy computations off-chain to save gas costs. We implemented the proposed protocol for the Ethereum smart contract platform and evaluated it regarding costs, performance, and memory to demonstrate its applicability.

In future work, we want to examine if there is potential to reduce the overhead of using smart contracts even further and integrate improvements from other protocols.
\section*{Acknowledgment}


The financial support from the Austrian Federal Ministry for Digital and Economic Affairs, the National Foundation for Research, Technology, and Development, and the Christian Doppler Research Association is gratefully acknowledged.

\bibliographystyle{ACM-Reference-Format.bst}
\balance
\bibliography{refs.bib} 

\end{document}